\documentclass[amsmath,amssymb,prl,hyperlink,twocolumn]{revtex4}

	\usepackage{graphicx}
	\usepackage{soul}
	\usepackage[colorlinks=true,citecolor=blue,linkcolor=magenta]{hyperref}
	\usepackage[usenames]{color}
	\usepackage{amsfonts}
	\usepackage{color}
	\usepackage{booktabs}
	\usepackage{multirow}
	\usepackage{float}
    \usepackage{times}

\begin{document}

\title{Efficient Measurement of Multiparticle Entanglement with Embedding Quantum Simulator}
\author{Ming-Cheng Chen$^{1,2 \dag}$, Dian Wu$^{1,2 \dag}$, Zu-En Su$^{1,2}$, Xin-Dong Cai$^{1,2}$, Xi-Lin Wang$^{1,2}$, Tao Yang$^{1,2}$,  Li Li$^{1,2}$, Nai-Le Liu$^{1,2}$, Chao-Yang Lu$^{1,2}$, and Jian-Wei Pan$^{1,2}$, \vspace{0.2cm}}

\affiliation{$^1$ Hefei National Laboratory for Physical Sciences at Microscale and Department of Modern Physics, University of Science and Technology of China, Hefei, Anhui 230026, China}
\affiliation{$^2$ CAS Centre for Excellence and Synergetic Innovation Centre in Quantum Information and Quantum Physics, University of Science and Technology of China, Hefei, Anhui 230026, China.}
\affiliation{$^{\dag}$ These authors contribute equally to this work.}
\date{\today}

\begin{abstract}
We reports direct and scalable measurement of multiparticle entanglement concurrence and three-tangle with embedding photonic quantum simulators. In this embedding framework [Phys. Rev. Lett. \textbf{111}, 240502 (2013)], $N$-qubit entanglement monotone, which associates with non-Hermitian operators, can be efficiently measured with only 2 (for even $N$) and 6 (for odd $N$) local measurement settings. Our experiment uses a multiphoton quantum simulator to mimic the dynamical entanglement evolution and track its concurrence and three-tangle.
\end{abstract}
\maketitle

\begin{figure*}[tb]
    \centering
        \includegraphics[width=0.9\textwidth]{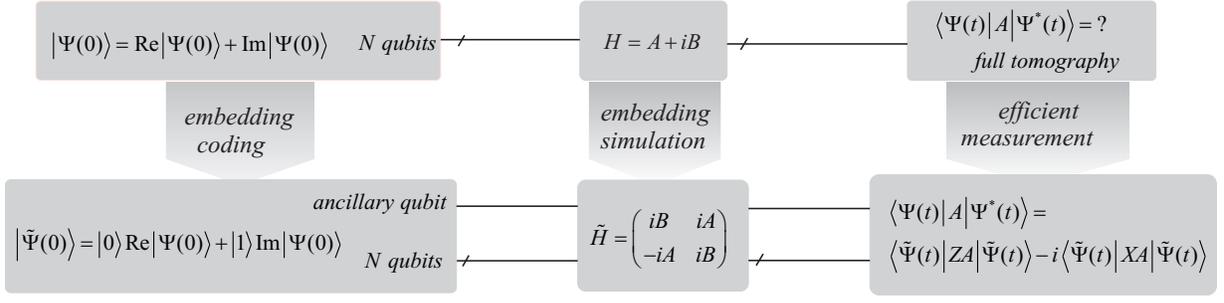}
\caption{Embedding simulator for efficient entanglement measure. The entanglement in original dynamic system $H$ cannot be directly measured as it contains complex terms like $\left\langle \Psi  \right|A\left| {{\Psi ^*}} \right\rangle $. Through embedding the initial state and Hamiltonian into an enlarged simulator assisted by an ancillary qubit, the complex terms are allowed to efficiently measure on the enlarged state. }
\label{fig1}
\end{figure*}

\begin{figure}[]
\centering
\includegraphics[width=0.52\textwidth]{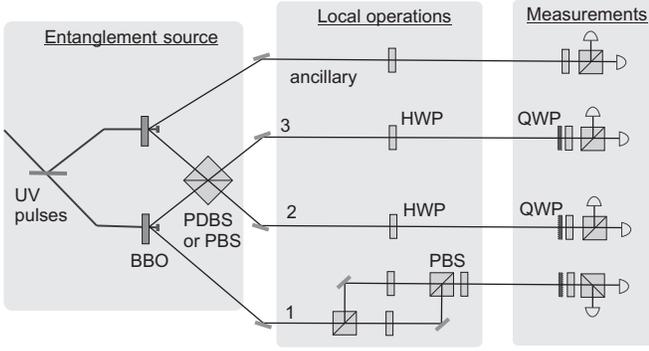}
\caption{Experimental setup. Ultraviolet (UV) femtosecond laser pulses (394 nm, 120 fs, 76 MHz) pass through two type-II BBO crystals to produce two pairs of entangled photons. Two single photons, one from each pairs, are mixed on a PDBS to generate W-type entangled state. To simulate the three-tangle system, the PDBS is switched to a PBS to generate GHZ-type entangled four photons. The multi-photon sources are sent to next stage to perform local operations by HWPs and a polarization dependent Mach-Zehnder interferometer. Finally, the photons are measured in by four-fold coincidence counting. BBO, beta-barium borate; PDBS, polarization dependent beam splitter with a transmission of $0.72$ for $H$ photons and $0.28$ for $V$ photons ; PBS, polarization beam splitter; HWP, half-wave plate; QWP, quarter-wave plate. $H$ and $V$ denote horizontal and vertical polarization.}
\label{fig2}
\end{figure}

\begin{figure*}[tb]
    \centering
        \includegraphics[width=0.8\textwidth]{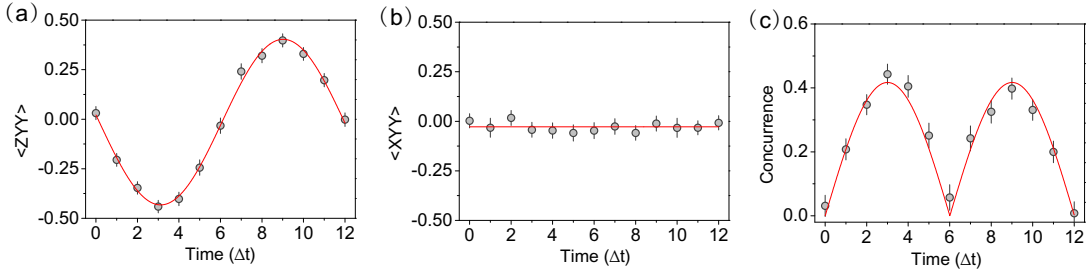}
\caption{Experimental results of entanglement dynamics in concurrence system. (a)(b) The time-dependent expectation of observable $ZYY$ and $XYY$. (c) The concurrence computing from two measurements of $ZYY$ and $XYY$. The error bars represent poissonian statistics in photon counting. The red curves are theoretical simulation counting for the imperfect interferometric visibility.}
\label{fig3}
\end{figure*}

\begin{figure*}[tb]
\centering
\includegraphics[width=1\textwidth]{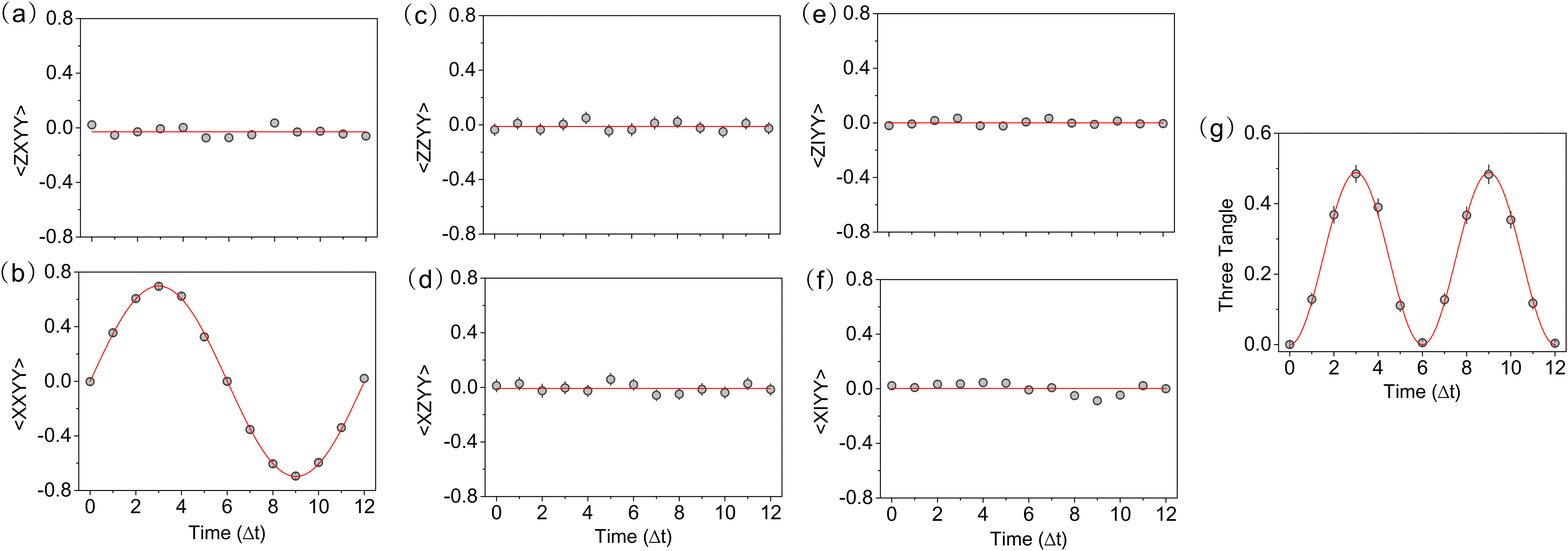}
\caption{Experimental results of entanglement dynamics in three-tangle system. (a)-(f) The time-dependent expectation of observable $ZXYY$, $XXYY$, $ZZYY$, $XZYY$, $ZIYY$, and $XIYY$. (g) The three-tangle computing from previous six measurements. The error bars represent poissonian statistics in photon counting. The red curves theoretical simulation counting for the imperfect interferometric visibility. }
\label{fig4}
\end{figure*}

Entanglement lies at the heart of quantum mechanics \cite{RMP2009Horodecki,RMP2012Pan} and plays an important role in quantum information sciences including quantum simulation \cite{RMP2014simulation}, quantum computation \cite{book2000Quantum-Computation,PRL2001One-way}, quantum network \cite{Nature1997teleportation,Nature2008Internet}, and quantum metrology \cite{Science2004metrology}. Immense experimental advance has allowed the creation of growing number of entangled particles in laboratories. To date, entanglement among fourteen trapped ions \cite{PRL2011fourteen-ion} , eight photons \cite{Nature2012eight-photon} , and five superconducting qubits \cite{Nature2014five-supperconduting} have been demonstrated.

Not only the generation of multiparticle entanglement, but also their verification and characterization, become exceedingly difficult for large number of quantum bits (qubits). Tracking the presence of multipartite entanglement has been considered as an important criteria for confirming quantumness in quantum machines \cite{dwave}. A fundamental obstacle for quantifying entanglement \cite{PRL1997quantifying-entanglement} is that its function usually involves a complex-conjugation operator, which doesn't associate with a Hermitian observable directly.

The conventional method of quantum state tomography \cite{PRL1999tomography} requires $\sim$$\,$$4^N$ measurements for $N$-qubit states, which is not scalable. Entanglement witness \cite{PhyRep2009detection, PRL2004witness,PRA2007witness} has been designed to detect the presence of certain classes of multiparticle entanglement with polynomial complexity of measurement settings. However, it is not universal---as different classes of entangled states would need different witnesses---and doesn't provide quantitative measures of entanglement. A single-shot measurement of concurrence \cite{Nature2006SingleMeasure} has been reported for a two-qubit state. Such methods, however, rely on collective measurements on multi-fold copy of the quantum states \cite{PRL2005GeneralizedConcurrence} which is experimentally demanding for multiparticle entanglement.


In this Letter, we report efficient measurement of concurrence \cite{PRL1998concurrence} and three-tangle \cite{PRA2000three-tangle} entanglement monotones \cite{PRA2005entanglement-monotone} evolution from initial separable states for two- and three-qubit system with embedded photonic quantum simulators. The photonic simulators are initialized in W- \cite{PRA2000W-state-three-qubit} and Greenberger-Horne-Zeilinger (GHZ) - \cite{1990GHZ} type entangled states, followed by local non-unitary operations,  to mimic the dynamics of multiparticle entanglement. With only 2 (for even number of qubits) and 6 (for odd number of qubits) measurements, we reconstruct the entanglement monotone dynamics, confirming the high efficiency of the embedding framework.

By definition, entanglement monotone \cite{PRA2005entanglement-monotone} is zero for separable states and doesn't increase in average under local operations and classical communication. The entanglement monotone for bi-particle system---concurrence---can be written as $\left| {\left\langle {{A_0}K} \right\rangle } \right|,$ where $A_0=YY $ (here, $X$, $Y$, and $Z$ are Pauli operators) and $K$ is the complex-conjugation operator ($ K\left| \psi  \right\rangle =\left| {\psi^*} \right\rangle $).
The monotone for tri-particle system---three-tangle---is $
\left| {{{\left\langle {{A_1}K} \right\rangle }^2} + {{\left\langle {{A_2}K} \right\rangle }^2} - {{\left\langle {{A_3}K} \right\rangle }^2}} \right|,
$ where $A_1=XYY$, $A_2=ZYY$, and $A_3=IYY$ ($I$ is the identify operator).
These definitions can be generalized to multiparticle case \cite{PRA2005entanglement-monotone}. For systems with even-$N$ particles, the monotone can be written in the form of concurrence, where $ {A_0}={Y^{ \otimes N}}$. For systems of odd-$N$ particles, the monotone is similar to three-tangle by replacing $A_1$, $A_2$, and $A_3$ with $ {A_1}={XY^{ \otimes N-1}}$, $ {A_2}={ZY^{ \otimes N-1}}$, and $ {A_3}={IY^{ \otimes N-1}}$. It can be noted that the entanglement monotone of multiparticle system can be expressed in the term of $ \left\langle {AK} \right\rangle $. Generally, $ \left\langle {AK} \right\rangle $ is not a real number as $AK$ is not a Hermitian observable.

Candia \textit{et al.} \cite{PRL2013embedding-simulator} proposed a protocol for efficient measurement of multipartite entanglement with embedding quantum simulators (see Fig.$\,$1).  In this embedding framework, the wavefunction of multiparticle system $\left| \psi  \right\rangle $ is separated into real and imagine components and entangled with an additional ancillary qubit, resulting in an enlarged state $\left| {\tilde \psi } \right\rangle $ as
\begin{align}
\left| {\tilde \psi} \right\rangle  = \left| 0 \right\rangle  \otimes {\mathop{\rm Re}\nolimits} \left| {\psi} \right\rangle  + \left| 1 \right\rangle  \otimes {\mathop{\rm Im}\nolimits} \left| {\psi} \right\rangle.
\end{align}
The real and imaginary parts of the value of $\left\langle {AK} \right\rangle $ can be directly measured by a Hermitian observable on the enlarged state by noting that
\begin{align}{\mathop{\rm Re}\nolimits} \left\langle {AK} \right\rangle  = \left\langle {ZA} \right\rangle,\\
{\mathop{\rm Im}\nolimits} \left\langle {AK} \right\rangle  =  - \left\langle {XA} \right\rangle,
\end{align}
where $Z$ and $X$ are the Pauli operators on the ancillary qubit. Therefore, for an even-$N$ particle system it will take only two measurements of observables $ZA_0$ and $XA_0$ on the enlarged state to evaluate the entanglement, and for an odd-$N$ particle system it needs only six measurements of observables $ZA_1$, $XA_1$, $ZA_2$, $XA_2$, $ZA_3$, and $XA_3$.

By embedding the initial wavefunction  $\left| {\psi (0)} \right\rangle$ and quantum dynamics $H$ into an enlarged simulator initialized in $\left| {\tilde \psi (0)} \right\rangle$ with an appropriate Hamiltonian $\tilde H$, the complex-conjugation expectation relation at various time still holds as \begin{align}\left\langle {AK} \right\rangle (t) = \left\langle {ZA} \right\rangle (t) - i\left\langle {XA} \right\rangle (t),\end{align} where \begin{align} H = A + iB, \tilde H = \left( {\begin{array}{*{20}{c}}
{iB}&{iA}\\
{ - iA}&{iB}
\end{array}} \right)\end{align} with real matrix  ${A^\dag } = A$ and ${B^\dag } =  - B$. This allow us to track the multiparticle entanglement dynamics efficiently.

Now we proceed with experimental demonstration using multi-photon quantum simulators \cite{simulators-review}.  To demonstrate the general working principle for multiparticle systems of even and odd number of qubits, we choose to implement the measurement of dynamical concurrence of a two-qubit entangled state and three-tangle of a three-qubit entangled state. The first example, a concurrence system, starts from an initial state $\left| {\psi (0)} \right\rangle  = \left| 0 \right\rangle  \otimes \left| 0 \right\rangle $ with a Hamiltonian $H = X \otimes Y + X \otimes Z $. The task is to evaluate the concurrence at arbitrary time $t$ associated with the unknown quantum state $\left| {\psi (t)} \right\rangle $. According to embedding framework, we set an enlarged embedding simulator with an initial state $\left| {\tilde \psi (0)} \right\rangle  = \left| 0 \right\rangle  \otimes \left| 0 \right\rangle  \otimes \left| 0 \right\rangle $ and a Hamiltonian $ \tilde H = I \otimes X \otimes Y - Y \otimes X \otimes Z $.

We implement this enlarged dynamical system in a three-photon compiled simulator. Our experimental setup is showed in Fig. 2.  We use single photons produced from spontaneous parametric down-conversion \cite{PRL1995type-II} as qubits, where the horizontal ($H$) and vertical ($V$) polarization are used to encode $ \left| 0 \right\rangle $ and $ \left| 1 \right\rangle $, respectively. The compiled simulator runs in three stages: (1) create an initial entangled source, (2) perform local operations, (3) finally, readout the expectation of observables $XA_0$ and $ZA_0$.

At stage (1), three single photons (ancillary, 1, and 2) are entangled in W-type state $(\left| {011} \right\rangle  - \left| {101} \right\rangle  + \left| {110} \right\rangle )/\sqrt 3 $ by post-selecting the photon 3 in $\left| 0 \right\rangle $ after mixing two pairs of Bell-state $(\left| {01} \right\rangle  + \left| {10} \right\rangle )/\sqrt 2 $ photons on a polarization dependent beam splitter (PDBS) with a transmission of 0.72 for $H$ photons and 0.28 for $V$ photons \cite{W}. At stage (2), the entangled photons are sent to perform local operations by half wave plates (HWPs) and a polarization dependent Mach-Zehner (MZ) interferometer. The ancillary photon was operated with $X$ operation and the photon 2 was operated with $Z$ and $X$ operations by passing through HWPs. Now the photons state are $(\left| {000} \right\rangle  + \left| {011} \right\rangle  - \left| {110} \right\rangle )/\sqrt 3 $. Then photon 1 passes through a polarization dependent  MZ interferometer operated with $\left( {\begin{array}{*{20}{c}}
{\cos \sqrt 2 t}&0\\
0&{\sin \sqrt 2 t/\sqrt 2 }
\end{array}} \right)$, where the coming photon is split into two spatial modes by a polarization beam splitter (PBS) regard to its polarization, and each pass through a HWP, then combine on a second PBS before anther HWP at the output port. After this stage, it will generate the desired enlarged state $\left| {\tilde \psi (t)} \right\rangle $, in the form of
$\cos (\sqrt 2 t)\left| {000} \right\rangle  + {\sin (\sqrt 2 t)} / {{\sqrt 2 }}\left| {011} \right\rangle  - {{\sin (\sqrt 2 t)}}/{{\sqrt 2 }}\left| {110} \right\rangle $.

We carry out the measurements on stage (3). By successive stepping the orientation of HWPs inside the MZ interferometer, the time parameter \textit{t}-dependent evolution is set at the points: $0$, $\Delta t$, $\Delta 2t$, $\Delta 3t$, $ \cdots $, where $\Delta t = \pi /12\sqrt 2 $. To evaluate the concurrence and reconstruct its dynamics, we made two measurements of $ZYY$ and $XYY$ at various time $t$. The experimental results are shown in Fig. 3(a) and (b). Then we computed the concurrence directly and shown it in Fig. 3(c). The concurrence dynamics is well fitted by $\left| {\alpha  \cdot \sin (\sqrt 2 t)/\sqrt 2 } \right|$ with $\alpha  = 0.59\pm0.01$, agreeing with the theoretical simulation scaled by an amplitude factor $\alpha $ counting for imperfect photonic interference visibilities in the creation of three-photon W state \cite{W}.

Next we present the second example, a three-tangle system with an initial state of $\left| {\psi (0)} \right\rangle  = \left| 0 \right\rangle  \otimes \left| 0 \right\rangle  \otimes \left| 0 \right\rangle $ and a Hamiltonian of $H = X \otimes X \otimes X$. To measure the entanglement of $\left| {\psi (t)} \right\rangle $, the enlarged state $\left| {\tilde \psi (t)} \right\rangle $ are implemented with an embedding simulator of initial state $\left| {\tilde \psi (0)} \right\rangle  = \left| {0000} \right\rangle $ with Hamiltonian $ \tilde H =  - YXXX $ .

The experiment was also run in three stages.
At stage (1), the initial entangled source $(\left| {0110} \right\rangle  + \left| {1001} \right\rangle )/\sqrt 2 $ is created by combining two pairs of Bell-state $(\left| {01} \right\rangle  + \left| {10} \right\rangle )/\sqrt 2 $ photons on a PBS. Then at stage (2), the photon 1 and 2 pass through a HWP to implement an $X$ operation. The photon 3 passes through a HWP to implement a $Z$ operation. And further, a non-unitary operator $\left( {\begin{array}{*{20}{c}}
{\cos t}&0\\
0&{\sin t}
\end{array}} \right)$ was applied to photon 1 by passing through a polarization dependent MZ interferometer. These two stages will make the desired enlarged state $\left| {\tilde \psi (t)} \right\rangle  = \cos (t)\left| {0000} \right\rangle  - \sin (t)\left| {1111} \right\rangle $.
At the final stage (3), we measured six observables $ZXYY$, $XXYY$, $ZZYY$, $XZYY$, $ZIYY$, and $XIYY$ at time $0$, $\Delta t$, $\Delta 2t$, $\Delta 3t$, $ \cdots $, where $\Delta t = \pi /12 $. The Fig4 (a)-(f) shows the experimental results. So we can directly compute the three-tangle. Fig. 4(g) show the evolution of three-tangle, theoretically fitted by ${\left| {\alpha  \cdot \sin (2t)} \right|^2}$ with $\alpha  = 0.70\pm0.01$, which is the visibility of the generated four-photon GHZ state in the superposition basis.

These two examples showed that the embedding framework can be used for efficient measurement of multiparticle entanglement dynamics. We have observed the dynamical entanglement oscillation well according with the theoretical prediction. The experimental imperfect performance of reducing the concurrence oscillation amplitude mainly comes from the high order emission of spontaneous parametric down-conversion photon pairs \cite{RMP2012Pan}.

In conclusion, we have presented a proof-of-principle demonstration of efficiently measuring entanglement dynamics for concurrence and three-tangle dynamics system on complied embedding photonic simulators. We have proved the high efficiency and feasibility of embedding framework for multipartite entanglement measurement. This paradigm can be applied to other physical systems such as trapped ions \cite{PRL2011fourteen-ion} and superconducting qubits \cite{Nature2014five-supperconduting} for scalable entanglement measurement, and may offer a promising way to study the entanglement of large-scale dynamical multipartite system. The embedding framework, exploiting additional qubits to reform the original wavefunction information, may inspire other exciting applications, such as simulation of non-unitary Majorana dynamics \cite{Majorana}.

\textit{Note}: When preparing our manuscript, we became aware of a related experiment work from Andrew White's group.

\textit{Acknowledgement}: This work was supported by the National Natural Science
Foundation of China, the Chinese Academy of Sciences and the National Fundamental
Research Program (grant no.2011CB921300).

\end{document}